
%
%
 \magnification 1200
 \baselineskip=20.10pt
 \def\rf{\hfill\break\noindent}
 
 \def\cl{\centerline}
 \def\ept{\epsilon_T}
 \centerline{\bf A COMMENT ON JUNCTION AND ENERGY CONDITIONS}
 \centerline{\bf IN THIN SHELLS}
 \vskip 1in
 \cl{Dalia S. Goldwirth\footnote*{dalia@wise7.tau.ac.il}}
 \cl{School of Physics and Astronomy, Raymond and Beverly Sackler,}
 \cl{Faculty of Exact Sciences, Tel-Aviv University, 69976}
 \cl{Tel-Aviv, Israel}
 \cl{J. Katz\footnote{$\dagger$}{jkatz@vms.huji.ac.il}}
 \cl{Racah Institute of Physics, Hebrew University, 91904 }
 \cl{Jerusalem, Israel}

 {\bf Abstract:}

 This comment contains a suggestion for a slight modification of
Israel's covariant formulation
of junction conditions between two spacetimes, placing both sides on
equal footing with normals having  uniquely defined orientations.
The signs of mass energy densities in thin shells at the junction
depend not only on the orientations of the normals and it is
useful therefore to discuss the signs separately. Calculations gain in clarity
by not choosing orientations in advance.
Simple examples illustrate our point and complete previous classifications
of spherical thin shells in spherically symmetric spacetimes relevant
to cosmology.
\par\vfil\eject
Two pieces of spacetimes may be glued together if their borders fit. That is a
junction condition. Having two spacetimes, $M$ and $\bar M$,
cut into two pieces $1$, $2$ and
$\bar 1$, $\bar 2$  with fitting borders, there are four possible assemblages:
$1 \bar 1$, $1 \bar 2$, $2 \bar 1$ and $2 \bar 2$.
To pick out one of the four amounts, in mathematical language, to chose two
pairs of signs or ``orientations". Discussions about orientations appear often
in the literature, but not always with the desired clarity.

We may want the glued pieces to connect smoothly across the border.
This is another junction condition which we will not be concerned with here.
We shall, on the contrary, look at  brutal changes of
curvature which are interpreted as ``thin material shells".
One expects ``physical" thin shells to have positive mass-energy densities.
This is an energy condition. Some or all of the four assemblies may satisfy
this energy condition. It is often a trivial matter to find out.
Figure 1 illustrates what we mean.

The most satisfying mathematical treatment of junction conditions is that
of Israel [1] in which there is no need for continuous coordinates. The
formalism does not treat, however, glued pieces on the same footing when it
comes to orientations.
Unit normal vectors to both pieces are said to  point in the same direction,
but a normal has two directions. It is not said which direction is taken.
Across thin spherical shells the direction is understandably taken from
smaller to larger radii, inside-outside.
But if, like in the case of spherical shells with two centers or shells
that are ``on the other side" of the Einstein-Rosen bridge,
the radii decrease on both sides
of the shell then the inside-outside language becomes inadequate if not
misleading.

Our aim is (a) to suggest a slight modification of Israel's formalism by
handling both sides of the junction on the same footing, (b)
to treat the energy conditions independently of the orientations
of the normals because energy conditions depend not only on the normals
and it does not always help to mix the two
and (c) to illustrate our point by treating spherically symmetric shells
in spherically symmetric spacetimes, completing previous classifications by
Sato [2], Berezin et.al. [3] and Sakai and Maeda [4]. We shall see,
 incidentally, how worthwhile it is not to chose
 arbitrary signs before there is a need for it.

We set $c=1$ and consider timelike shells. A few remarks are made on
lightlike shells at the end.

Take one of the 4 assemblages in figure 1, say,  $1 \bar 1$. Each piece has a
metric $ds^2 = g_{\mu \nu}dx^{\mu}dx^{\nu}$ and
$d \bar s^2 = \bar g_{\mu \nu}d \bar x^{\mu}d \bar x^{\nu}$
(with signature -2) described in its own most convenient local system
of coordinates $x^{\lambda}$ and $\bar x^{\lambda}$.
The common spacelike hypersurface $\Sigma$ is described by two sets of
three equations $x^{\lambda}(\theta^a)$ and $\bar x^{\lambda}(\bar {\theta}^a)$
with convenient parameterizations
$\theta^a$ and $\bar {\theta}^a$ (a,b,c,d=0,2,3). The metric of $\Sigma$
as viewed in 1 is
$$
ds_{\Sigma}^2 = g_{\mu \nu} {\partial x^{\mu} \over \partial \theta^a}
{\partial x^{\nu} \over \partial \theta^b} d\theta^a d \theta^b
=  g_{\mu \nu}h^{\mu}_a h^{\nu}_b d\theta^a d \theta^b =
\gamma_{ab}d\theta^a d \theta^b \ \ \  \eqno (1)
$$
and has the signature -1; as viewed in $\bar 1$ it is
$$
ds_{\Sigma}^2 = \bar g_{\mu \nu} {\partial \bar x^{\mu} \over \partial \bar
{\theta}^a}
{\partial \bar x^{\nu} \over \partial \bar{\theta}^b} d\bar{\theta}^a
d \bar{\theta}^b
=  \bar g_{\mu \nu} \bar h^{\mu}_a \bar h^{\nu}_b d\bar{\theta}^a
d \bar{\theta}^b =  \bar{\gamma}_{ab} d\bar{\theta}^a
d \bar{\theta}^b \ \ \ . \eqno (2)
$$
There must exist a transformation of the local coordinates $\theta^a$
and $\bar {\theta}^b$ such that
$$
\gamma_{ab} = {\partial \bar {\theta}^c \over \partial \theta^a}
{\partial \bar {\theta}^d \over \partial \theta^b} \bar {\gamma}_{cd}
\ \ \ . \eqno (3)
$$
Equations (3) are the junction conditions between $1$ and $\bar 1$
along $\Sigma$.

The components of the unit normal (spacelike) vectors to $\Sigma$,
$n^{\lambda}$ in $1$ and $\bar n^{\lambda}$ in $\bar 1$, satisfy the following
pair of four equations
$$
n_{\lambda}n^{\lambda} = -1 \ \ , \ \ n_{\lambda}h^{\lambda}_a = 0;
$$
$$
\bar n_{\lambda} \bar n^{\lambda} = -1 \ \ ,
\ \ \bar n_{\lambda} \bar h^{\lambda}_a = 0 \ \ \ . \eqno (4)
$$
These equations do not define the orientations of the normals, i.e.
their signs. To that effect consider two small vectors, $\delta x^{\lambda}$
in $1$ and $\delta \bar x^{\lambda}$ in $\bar 1$. For $n^{\lambda}$ and
$\bar n^{\lambda}$ {\it to be directed in their own spaces}, as shown in
 figure 1, the following
inequalities must hold
$$
n_{\lambda} \delta x^{\lambda} < 0 \ \ ,\ \ \bar n_{\lambda} \delta
\bar x^{\lambda} < 0 \ \ \ . \eqno (5)
$$
These inequalities define the orientations.

Having written the junction conditions - equation (3) - and defined the
orientations - equation (5) - we calculate the energy tensor $\tau_{ab}$
of the thin shell. It is obtained from $\Sigma$'s two external
curvature tensors components $K_{ab}$ in $1$ and $\bar K_{ab}$
in $\bar 1$. Following Eisenhart [5]
$$
K_{ab} = -h^{\mu}_a h^{\nu}_b D_{\mu}n_{\nu} \ \ , \ \
\bar K_{ab} = -\bar h^{\mu}_a \bar h^{\nu}_b \bar D_{\mu} \bar n_{\nu}
 \ \ . \eqno (6)
$$
With these K's one constructs ``Lanczos tensors" whose components
$L^b_a$, $\bar L^b_a$ are
$$
L^b_a = \delta^b_a K - K^b_a  \ \ , \ \
\bar L^b_a = \delta^b_a \bar K - \bar K^b_a \eqno (7)
$$
where $ K^b_a = \gamma^{bc}K_{ca} $, $K = K^a_a$ and the like for
barred $K$'s and  barred $\gamma$'s. The Gauss - Codazzi - Mainardi
identity relates the Lanczos tensors to the energy momentum tensors
of each side
$$
\nabla_b L^b_a = 8\pi G n_{\mu}T^{\mu}_{\nu}h^{\nu}_a \ \ \ \
\bar \nabla_b \bar L^b_a = 8\pi G \bar n_{\mu} \bar T^{\mu}_{\nu} \bar
h^{\nu}_a
\eqno (8)
$$
$\nabla_b$ denotes a covariant derivative on $\Sigma$. The energy tensor of the
shell is then given by the sum of $L$'s
$$
8 \pi G \tau^b_a = L^b_a + \bar L^b_a
 \ \ . \eqno (9)
$$
Note the (+) sign instead of the usual (-) sign in accordance with our choice
of orientations. (See Israel [1] for an heuristic justification of $\tau^b_a$.)
The equations of motion of the shell follow from the sum of the
two equations (8),
together with equations (9)
$$
\nabla_b \tau^b_a = n_{\mu}T^{\mu}_{\nu}h^{\nu}_a +
\bar n_{\mu} \bar T^{\mu}_{\nu} \bar h^{\nu}_a
 \ \ , \eqno (10)
$$
to which we should add, in general,  an equation of state for completeness.

There are various energy conditions in Hawking and Ellis [6].
 We adopt the dominant energy condition which implies positive mass-energy
density, flows of energy not faster than light and an upper bound on pressures
as well as on tensions.
In a diagonal representation
of $\tau^b_a \equiv$ diag($\tau^0_0$, $\tau^2_2$, $\tau^3_3$), we must have
$$
\tau^0_0 \equiv \sigma > 0  \ \ , \eqno (11)
$$
and
$$
 with\   \ \ \tau^2_2=\tau^3_3 \equiv -\Pi,\ \ \ \ \ \
   \mid \Pi \mid < \sigma \ \ .  \eqno (12)
$$
These are the energy conditions referred to in the introduction.
We turn now to two examples of spherical shells in spherically symmetric
spacetimes that illustrate our purpose and complete known classifications.
The key equations are (3), (5), (9), (11) and (12).

   {\it (i) Spherical shells in static spherically symmetric spacetimes
 on both sides with metrics of the form}
$$
ds^2 = a(r)dt^2 - b(r)dr^2 - r^2 d\Omega^2 ,  \ \ \ \ \ \
d\Omega^2 \equiv d\theta^2 + sin^2\theta d\phi^2  \ \ . \eqno (13)
$$

Let $t=T(\tau)$, $r=R(\tau)$ be the equations of the spherical shell in $1$,
where $\tau$ represents the proper time on $\Sigma$ and let
$\bar t = \bar T(\tau)$, $\bar r= \bar R(\tau)$ be the equations of
$\Sigma$ in $\bar 1$. The junction conditions, equation (3),  of $1$
and $\bar 1$ give first $\bar R = R$. The metric of $\Sigma$ in
coordinates $\theta^a = \bar \theta^a = (\tau, \theta, \phi)$ is
$$
ds_{\Sigma}^2 = d\tau^2 - R^2(\tau)d\Omega^2 \ \ . \eqno (14)
$$
Second, $T(\tau)$ and $\bar T(\tau)$ must be related to $R(\tau)$ by the
requirement
that $\tau$ is the proper time on $\Sigma$:
$$
\dot T = sign(\dot T) \sqrt{A^{-1}(1 + B \dot R^2)} \ \ \ \
\dot {\bar T} = sign(\dot {\bar T})
\sqrt{\bar A^{-1}(1 + \bar B \dot R^2)} \ \ , \eqno (15)
$$
where
$$
 A=a(R(\tau)) \ \,\ \ B=b(R(\tau)) \ \,
\ \ \bar A = \bar a(R(\tau)) \ \,\ \
\bar B = \bar b(R(\tau)) \ \ . \eqno (16)
$$
Dots represent derivatives with respect to $\tau$.
It is convenient to define the future on $\Sigma$ in the direction
of increasing $\tau$.
Sign($\dot T$) and  sign($\dot {\bar T}$)
is one of the  pair of orientations we mentioned in the beginning.
The other pair is that of the unit normals vectors on $\Sigma$; here
$$
n_{\lambda} = sign(n)\sqrt{AB}(\dot R, -\dot T, 0, 0) \ \ \ \
\bar n_{\lambda} = sign(\bar n)\sqrt{\bar A \bar B}
(\dot  R, -\dot {\bar T}, 0, 0) \ \ . \eqno (17)
$$
All unspecified signs introduced so far must satisfy equation (5) for all
vectors on $\Sigma$, $\delta x^{\lambda} = (\delta t, \delta r, 0 ,0)$ in
$1$ and  $\delta \bar  x^{\lambda} = (\delta \bar t, \delta \bar r, 0, 0)$
in $\bar 1$. The signs of $\delta t$, $\delta r$, $\delta \bar t$ and
$\delta \bar r$ are all known once we have laid out coordinates.

Consider first non static shells ($\dot R \ne 0$). Equation (5) for pure
timelike vectors ($\delta r = \delta \bar r = 0$) give
$$
sign (n) = - sign(\delta t)sign(\dot R) \ \ \ \
sign (\bar n) = - sign(\delta \bar t)sign(\dot R)
 \ \ . \eqno (18)
$$
For pure spacelike vectors ($\delta t = \delta \bar t = 0$) one obtain
from equation (5)
$$
sign(\dot T) = sign(\delta r)sign(n) \ \ \ \
sign(\dot {\bar T}) = sign(\delta \bar r)sign(\bar n)
 \ \ . \eqno (19)
$$
Thus for moving shells, orientations are completely defined.

For static shells ($\dot R = 0$), pure timelike vectors provide no
equations (0=0); only spacelike vectors provide equations (19). One  usually
 take $\dot T$ and $\dot{\bar T}$ in the direction of
$\tau$ but this is a matter of convenience.

Whether $\dot R$ is zero or not, we can calculate the mass energy
tensor $\sigma$ from equation (9); here $\sigma$ is given by
$$
\sigma = -{1 \over 4 \pi G R}\left [sign(\delta r) sign(B)
\sqrt{ B^{-1} +  \dot R^2} + sign(\delta \bar r) sign(\bar B)
 \sqrt{\bar B^{-1} +  \dot R^2} \right ]
\ \ . \eqno (20)
$$
{}From equation (10) one obtain an expression for
 pressures ($\Pi > 0$) or  tensions ($\Pi < 0$) in the shell
$$
\Pi = -{d (\sigma R^2) \over dR^2}
+ {R \over 2}\left [sign(n)\dot T(\rho + P) + sign(\bar n) \dot{\bar T}
(\bar {\rho} + \bar P) \right ] \ \ , \eqno (21)
$$
 $\rho$, $P$ and $\bar {\rho}$, $\bar P$ are the mass energy densities
and pressures or tensions on both sides of $\Sigma$. With an equation of state
$\Pi(\sigma)$ equation (21) becomes a first order differential equation
for R. If we can calculate $R(\tau )$ with appropriate initial conditions we
 can work our way back and calculate all other time dependent quantities:
$T(\tau), \bar T(\tau), \sigma(\tau)$ and $\Pi(\tau)$.

We turn now to the energy condition, equation (11). There is a distinction
to be made between spacetimes with equal $B$'s and with different $B$'s.
If $B \ne \bar B$ call bar the side with the smallest $B$:
$\bar B < B$. Then $\sigma > 0$ if
$$
sign(\delta \bar r)  sign(\bar B) = - 1\ \ \ \ (\bar B < B)
\ \ . \eqno (22)
$$
For instance if the bar space is flat ($\bar B = 1$) it must
have a center ($\delta \bar r < 0$). Equation (22) leaves two acceptable
options  among the 4 possible ones shown in figure 1,
$$
\sigma = {1 \over 4 \pi G R}\left [\sqrt{\bar B^{-1} +  \dot R^2} -
sign(\delta r) sign(B)\sqrt{B^{-1} +  \dot R^2} \right )
\ \ . \eqno (23)
$$
Whatever $sign(B)$ is, the non bared side may be ``centered"
($\delta r < 0$) or not.  If it is flat and centered,
our shell is in a closed universe with two centers like
in the example given by Lynden-Bell et al. [7]. In normal cases, there is
one center only.

If $\bar B = B$, the situation is quite different since
$$
\sigma = -{1 \over 4 \pi G R}sign(B)\left [ sign (\delta r)+
sign(\delta \bar r) \right ] \sqrt{B^{-1} + \dot R^2}
\ \ . \eqno (24)
$$
There is only one admissible assemblage in this case
$$
sign(\delta r) = sign(\delta \bar r) = -sign (B)
\eqno (25)
$$
for which
$$
\sigma = {1 \over 2 \pi G R}\sqrt{B^{-1} + \dot R^2}
\ \ . \eqno (26)
$$
If we represent the mass energy of the shell $4 \pi R^2 \sigma$ by $M_s$
, $\Pi$ can be written as
$$
\Pi = -{d(\sigma R^2) \over dR^2} - {1 \over 2}GM_s(\rho + P)B
\ \ . \eqno (27)
$$
Thus two centered static shells ($\dot R = 0$) are the only admissible
assemblages in flat spacetimes ($B=1 \ ,\ \ \rho=P=0$).
They satisfy the energy condition (12)
as is obvious from the fact that now
$$
\sigma = {1 \over 2 \pi G R} \ \ and  \ \ \
\Pi = - {1 \over 2}\sigma
\ \ . \eqno (28)
$$
The class of two centered shells in flat spaces completes Sato's
classification [2].

 Equation (28) represents a rather exotic
type of shells. With $R = 1 A.U.$,
the mass energy density must be $\sigma = 2.5~10^{14} gr~~~ cm^{-2}$ and the
tension $\Pi$ is as high as $10^{35}~ dyne~~~ cm^{-1}$.

{\it (ii) Spherical shells in Robertson-Walker spacetimes on both sides
 with metrics of the form}
$$
ds^2 = dt^2 -a^2(t)\left[ {dl^2 \over 1 - k l^2} + l^2d\Omega^2 \right]
\ \ \ \   k=0, \pm 1 \ \ .  \eqno (29)
$$

Useful quantities here are the Hubble ``constants"
$$
H = {1 \over a} {da \over dt} \ \ , \ \ \bar H = {1 \over \bar a}
{d \bar a \over d \bar t} \eqno (30)
$$
and their relation to the mass energy densities on both sides (Einstein's 0-0
equation)
$$
H^2 + {k \over a^2} = {8 \pi G \rho \over 3} \ \ \ \
\bar H^2 + {\bar{k} \over \bar a^2} = {8 \pi G \bar {\rho} \over 3}
\ \ .  \eqno (31)
$$
Let $t=T(\tau)$ and $l=L(\tau)$ be the equations of the shell $\Sigma$
in $1$ and $\bar t = \bar T(\tau)$, $\bar l = \bar L(\tau)$ in $\bar 1$.
{}From equations (3) we find that the junction conditions are here
$$
a(T(\tau))L(\tau) = AL = \bar a(\bar T(\tau))\bar L(\tau) = \bar A \bar L =
R(\tau) \ \ . \eqno (32)
$$
The metric $ds_{\Sigma}^2$ of $\Sigma$
is the same as equation (14) for the previous class of metrics and so is the
parameterization ${\theta}^a$.
Moreover, $T$ and $\bar T$ are related to $R(\tau)$ by the condition that
 $\tau$ is
the proper time on the shell, seen from side $1$ or side $\bar 1$:
$$
\dot T^2 - {A^2 \over 1-k L^2}\dot L^2=1 \ \ \ \
\dot {\bar T}^2 - {\bar A^2 \over 1-\bar k \bar L^2}
\dot {\bar L}^2 = 1
\ \ . \eqno (33)
$$
Following equation (32)
$$
A \dot L = \dot R - RH\dot T \ \ \,\ \ \
\bar A \dot {\bar L} = \dot R - R\bar H \dot {\bar T}
 \eqno (34)
$$
using equation (33) this leads to the following explicit equation for $\dot T$
$$
\dot T = {1 \over 1 - {8\pi G \rho \over 3} R^2} \left [ -R\dot R H
+ \ept \sqrt{ (1-k L^2)(1 + \dot R^2 - {8 \pi G \rho \over 3}
R^2) } \right ]
\ \ , \eqno (35)
$$
where $\ept = \pm 1$.
The same equation holds for $\dot {\bar T}$ with bars. As before, we
take the future of $\Sigma$ in the $d\tau > 0$ direction. The signs
of $\dot T$ and $\dot {\bar T}$ may be fixed by equations (5).

Other undefined signs appear in the unit normal vectors on both sides of
$\Sigma$
$$
n_{\lambda} = sign(n) {A \over \sqrt{1 - k L^2} } (\dot L, -\dot T, 0, 0)
\ \ \ \ \bar n_{\lambda} = sign(\bar n) {\bar A \over
\sqrt{1 - \bar k \bar L^2} } (\dot{\bar L}, -\dot{\bar T}, 0, 0)
\ \ . \eqno (36)
$$
As for the orientations: if $\dot L$ and $\dot {\bar L}$ are both non zero,
equations (5) yield, for pure timelike vectors $(\delta l = \delta \bar{l}=0)$,
$$
sign(n) = -sign(\delta t)sign(\dot L) \ \ \ \
sign(\bar n) = -sign(\delta \bar t) sign(\dot {\bar L})
\eqno (37)
$$
and for pure spacelike vectors $(\delta t = \delta \bar{t}=0)$
$$
sign(\dot T) = sign(\delta l)sign(n) \ \ \ \
sign(\dot {\bar T}) = sign(\delta \bar l) sign(\bar n)
\ \ .\eqno (38)
$$
If both $\dot L$ and $\dot {\bar L}$ are zero, we have only equation (38) to
fix orientations. It is a matter of convenience how we chose then
$sign(\dot T)$ and $sign(\dot {\bar T})$.
In a mixed situation, i.e. if either $\dot L$ or $\dot {\bar L}$ is
equal to zero, the orientations are given by  equations (38)and one of
equations (37) with $sign(\dot T)$ or $sign(\dot {\bar T})$ undefined.
The mass energy density defined by equation (9) can be
written here as
$$
\sigma = -{1 \over 4\pi G R} \left[  sign (n)\epsilon_T
\sqrt {1 + \dot R^2 - {8 \pi G \rho \over 3}R^2} + sign ( \bar n) \epsilon_{
\bar T }
\sqrt {1 + \dot R^2 - {8 \pi G \bar {\rho} \over 3} R^2} \right]
\eqno (39)
$$
which is similar to equation (20). {}From equation (10) one obtains then
$$
\Pi = -{d(\sigma R^2) \over dR^2} + {R \over 2 \dot R}\left [sign(n)
{A\dot L \dot T \over \sqrt{1-k L^2}}(\rho + P) + sign(\bar n)
{\bar A \dot{\bar L} \dot{\bar T} \over \sqrt{1-\bar k \bar L^2}}(\bar
{\rho} + \bar P) \right]
 \eqno (40)
$$
provided $\dot R \ne 0$.
If $\dot R = 0$, one can show that both $\rho + P$ and $\bar {\rho} + \bar P$
must be zero and that the second term on the right hand side of equation
(40) vanishes.

The energy condition $\sigma > 0$ is
similar to that of the previous example;
it depends on whether the $\rho$'s on both sides are equal or not.
If $\rho \ne \bar {\rho}$ let the bar denote the least dense side,
$\bar {\rho} < \rho$. In this case the requirement that $\sigma > 0$
yields (see equation (39))
$$
sign(\bar n) \epsilon_{\bar T} = -1
\eqno (41)
$$
and then
$$
\sigma = {1 \over 4\pi G R} \left[\sqrt {1 + \dot R^2 -
{8 \pi G \bar{\rho} \over 3} R^2} - sign (n)\epsilon_T
\sqrt {1 + \dot R^2 - {8 \pi G \rho \over 3}R^2}  \right ]
\ \ . \eqno (42)
$$
There are again two distinct possible assemblages in this case for
$sign (n)\epsilon_T =1$ and  $sign (n)\epsilon_T=-1$.
Classifications of different cases with $\rho \ne \bar {\rho}$ have
been given by Sakai and Maeda [4].
If $\bar {\rho} = \rho$, however,
$$
\sigma = -{1 \over 4\pi G R} \left[ sign (n)\epsilon_T + sign (\bar
n)\epsilon_{\bar{T}}
\right] \sqrt {1 + \dot R^2 - {8 \pi G \rho \over 3} R^2}
\ \ . \eqno (43)
$$
The only admissible orientations are those in which
$$
sign (n)\epsilon_T = sign (\bar n)\epsilon_{\bar T} = -1
\eqno (44)
$$
for which
$$
\sigma = {1 \over 2\pi G R} \sqrt {1 + \dot R^2 - {8 \pi G \rho
 \over 3} R^2}
\ \ . \eqno (45)
$$

As a simple example consider shells with static surfaces ($\dot R = 0$)
in spatially flat ($k = 0$) expanding ($H = \sqrt{8 \pi G \rho / 3}$)
de Sitter spacetimes ($\rho + P = 0$).
In this case, the solution of equation (35) with $\tau_0 = T_0 = 0$ is
$$
T = sign(\dot T){\tau \over \sqrt{1-(HR)^2} }
\eqno (46)
$$
and
$$
L = R~~e^{-HT}
\ \ . \eqno (47)
$$
{}From equations (46) and (47) and from equation (35) which tell us that
$sign(\dot T)=\epsilon_T$ we see that
$$
-sign(\dot{L}) = sign(\dot T)= \epsilon_T
\ \ . \eqno (48)
$$
Orientations are given by equations (37) and (38) which, combined
 with equation (48),
imply
$$
sign(\delta t) = sign(\delta l)= \epsilon_T sign(n)
\ \ . \eqno (49)
$$
Equations (46) to (49) hold naturally in the bar space as well.
The mass energy density is given by equation (43) which for this case
gets the simple form
$$
\sigma = -{1 \over 4\pi G R} \left[ sign (\delta l) + sign (\delta \bar l)
\right ] \sqrt {1 - (HR)^2}
\ \ . \eqno (50)
$$
The energy condition (11): $\sigma > 0$ admits only two centered shells
$$
sign (\delta l) = sign (\delta \bar l) = -1
\ \ . \eqno (51)
$$
This fixes other orientations given in equations (48) and (49). With (51),
$$
\sigma = {1 \over 2 \pi G R} \sqrt {1 - (HR)^2}
\ \ . \eqno (52)
$$
{}From equations (40) and (52) one obtains
$$
\Pi = {1 \over 4\pi G R}{ 1 - 2(HR)^2 \over \sqrt {1 - (HR)^2} }
\ \ . \eqno (53)
$$
Energy condition (12): the shell is under pressure if $(HR)^2<1/2$; it may be
in tension provided $1/2 < (HR)^2 < 3/4$.

We have not treated lightlike shells, but this can easily be done along the
same
lines. Here, unit normal vectors are inside $\Sigma$ and new
(nul) vectors $N^{\lambda}$ and $\bar N^{\lambda}$ are needed
to describe the motion of the shell (see Barrab{\` e}s and Israel [8]).
The new vectors must point out of $\Sigma$. As such, we would orient
these $N$'s in their own space like we have done with the n's
(see equations (5))
$$
N_{\lambda} \delta x^{\lambda} < 0 \ \ , \ \
\bar N_{\lambda} \delta \bar x^{\lambda} < 0
\ \ . \eqno (54)
$$
We shall not elaborate further on lightlike shells.
\par\vfil\eject
\cl {\bf REFERENCES}
\rf
[1] Israel W 1966 {\it Nuovo Cimento} B {\bf 44} 1; 1967
{\it Nuovo Cimento} B {\bf 48} 463
\rf
[2] Sato H 1986 {\it Prog. Theor. Phys.} {\bf 76} 1250
\rf
[3] Berezin V A, Kuzmin V A and Tkachev I I 1987 {\it Phys. Rev.} D {\bf 36}
2919
\rf
[4] Sakai N and Maeda K 1993 gr-qc/9311024
\rf
[5] Eisenhart C 1949 {\it Riemannian Geometry} Princeton University
Press
\rf
[6] Hawking S W and Ellis G F R  1973 {\it The Large Scale Structure
of Space-Time} Cambridge University Press
\rf
[7] Lynden-Bell D , Katz J and Redmount I H  1989 {\it MNRAS} {\bf 239} 201
\rf
[8] Barrab{\` e}s C and Israel W 1991 {\it Phys. Rev.} D {\bf 43} 1129
$$
$$
\cl {\bf FIGURE}
Fig. 1. A two dimensional example: fitting  a plane to a cone along
a circle. The pieces are numbered in the top figure (a). The four
configurations
in (b) to (e) show how the unit normal vectors are oriented.
In this example, there must be a shell along the junction.
\bye